\documentclass[%
 reprint,
 amsmath,amssymb,
 aps,
pra,
floatfix,
]{revtex4-2}

\usepackage{graphicx}% Include figure files
\usepackage{dcolumn}% Align table columns on decimal point
\usepackage{bm}% bold math
\usepackage[hidelinks]{hyperref}% add hypertext capabilities
\usepackage[mathlines]{lineno}% Enable numbering of text and display math
\usepackage{amsmath}
\usepackage{amssymb}
\usepackage{amsfonts}
\usepackage{array}
\usepackage{braket}
\usepackage{booktabs}
\usepackage{comment}
\usepackage{float}
\usepackage{graphicx}
\usepackage{array}
\usepackage{tikz}
\usetikzlibrary{decorations.pathreplacing}
\usetikzlibrary{positioning}
\usetikzlibrary{calc}
\usepackage{subcaption}

\begin{document}

\preprint{APS/123-QED}

\title{Physics-inspired neural networks as quasi inverse of quantum channels}% Force line breaks with \\

\author{Sameen Aziz}
%Lines break automatically or can be forced with \\
\author{Muhammad Faryad}%
%\href{mailto:muhammad.faryad@lums.edu.pk}
\email{muhammad.faryad@lums.edu.pk}
\affiliation{Department of Physics, Lahore University of Management Sciences (LUMS), Opposite Sector U, D.H.A, Lahore 54792, Pakistan}%

\date{\today}

\begin{abstract}
Quantum channels are not invertible in general. A quasi-inverse allows for a partial recovery of the input state, but its analytical results are found only in a restricted space of its parameters. This work explores the potential of neural networks to find the quasi-inverse of qubit channels for any values of the channel parameters while keeping the quasi-inverse as a physically realizable quantum operation. We introduce a physics-inspired loss function based on the mean of the square of the modified trace distance (MSMTD). The scaled trace distance is used to so that the neural network does not increase the length of the Bloch vector of the quantum states, which ensures that the network behaves as a completely positive and trace-preserving (CPTP) quantum channel. The Kraus operators of the quasi-inverse channel were obtained by performing the quantum process tomography on the trained neural network.
\end{abstract}

\maketitle

%\tableofcontents

\section{Introduction}
A quantum channel, like its classical counterpart, transmits information from one place to another. However, these transmissions are rarely ever perfect, and the system is often affected by noise. One of the central areas of research in quantum information is to alleviate the effect of noise and reduce errors~\cite{NC2010}. If the effect of a channel can somehow be reversed, then it may be possible to recover the input state. Quantum channels, however, cannot be inverted unless they are simple unitary channels. Therefore, an approximate inverse known as a quasi-inverse was proposed~\cite{quasi-inverse} {with the idea being that it increased the average fidelity of the channel for pure input states. This idea was extended for mixed input states using the trace distance in Ref.~\cite{faizan-2025}.} General analytical expressions of quasi-inverse have only been found for qubit channels but not for higher-dimensional channels~\cite{Shahbeigi_2021}. However, the quasi-inverse of qubit channels also exists only in a small range of parameters of the quantum channels \cite{quasi-inverse}.

Machine learning has been an active area of research for the past few decades, and several domains of science have benefited from it. Neural networks, in particular, have gained much popularity owing to their ability to learn complex patterns. Research has demonstrated the usefulness of neural networks in various areas of physics. For example, physics-informed neural networks designed specifically to solve physics-related tasks were shown to be very effective~\cite{RAISSI2019686}. Several studies show how the integration of machine learning and quantum information could provide computational advantages~\cite{yamasaki2023, Xiao_2023}. Neural networks, also called quantum neural networks, have been successfully used for learning an unknown unitary  ~\cite{Beer_2020} as well as to reconstruct different quantum channels~\cite{Ma_2023}. More complex neural networks, such as quantum autoencoders, have been used to successfully denoise the Greenberger-Horne-Zeilinger (GHZ) state~\cite{Bondarenko_2020}. The robustness of quantum autoencoders in quantum technologies has also been demonstrated~\cite{Mok_2023}. Quantum neural networks can be used for error-correction purposes~\cite{Chalkiadakis23, zhong2024}. Neural networks have also been used to recover noise-free states from noisy channel outputs~\cite{Morgillo2024}.

In this paper, we use neural networks to find the quasi-inverses of various qubit channels.
Although the quasi-inverse of qubit channels is found analytically, it only exists under restricted conditions on the parameters of the quantum channels. For example, the analytical quasi-inverse exists only if the error probability of Pauli error channels is sufficiently high~\cite{quasi-inverse}. However, in many quantum technologies, including quantum computing, error occurs with small probability, and an analytical quasi-inverse cannot be used. So, we set out to approximate inverses for any quantum channel using neural networks as a general function approximator. We constrained the loss function to make the neural network act as trace-preserving quantum channels. The plan of the paper is as follows: In Sec. \ref{model}, we describe the problem setup and loss function according to the physics of the problem. The training data creation and the training of the neural network are presented in Sec. \ref{training}. Section \ref{tomography} contains the numerical results for a few simple channels and the results of the quantum process tomography on the trained neural networks. Finally, the concluding remarks are presented in Sec. \ref{conc}.

\section{Physics inspired loss function}\label{model}
A density matrix of a single-qubit state has the following form
\begin{align}
   \rho &=\frac{1}{2}(I+\textbf{r}\cdot \boldsymbol{\sigma}),
\end{align}
where $I$ is the identity matrix, $\textbf{r}$ is the Bloch vector of the state represented by $\rho$ and $\boldsymbol{\sigma}=(\sigma_1,\sigma_2,\sigma_3)$ is the vector of the Pauli matrices.

The bit flip channel is
\begin{align}
        \mathcal{E} (\rho) &= (1-p)\rho + p(\sigma_{1}\rho \sigma_{1}),
\end{align}
where $p \in [0,1]$ is the error probability.

The bit-phase flip channel is defined as
\begin{align}
        \mathcal{E} (\rho) &= (1-p)\rho + p(\sigma_{2}\rho \sigma_{2}).
\end{align}

The phase flip channel is defined as
\begin{align}
        \mathcal{E} (\rho) &= (1-p)\rho + p(\sigma_{3}\rho \sigma_{3}).
\end{align}

The amplitude damping channel is defined as
\begin{align}
        \mathcal{E} (\rho) &= E_{0}\rho E_{0}^{\dag} + E_{1}\rho E_{1}^{\dag},
\end{align}
where $E_{0} = \begin{pmatrix}
    1 & 0\\
    0 & \sqrt{1-\gamma}\\
\end{pmatrix}$ and $E_{1} = \begin{pmatrix}
    0 & \sqrt{\gamma}\\
    0 & 0\\
\end{pmatrix}$ and $\gamma \in [0,1]$ is the damping parameter.

\begin{figure}[H]
\centering
    \begin{tikzpicture}[
    roundnode/.style={circle, draw=black!60, fill=gray!10, very thick, minimum size=7mm},
    squarednode/.style={rectangle, draw=black!60,fill=gray!10,very thick, minimum size=5mm}
    ]
    \node[roundnode](rho) {$\textbf r$};
    \node[squarednode](qc)[right =2cm of rho]{Quantum channel};
   \node[roundnode] (qc_out)[ right= 2cm of qc] {$\textbf{r}^{\,\prime}$};
    \draw[->] (rho.east) -- (qc.west);
    \draw[->] (qc.east) -- (qc_out.west);
    \node[roundnode](qc_out2)[below = of rho] {$\textbf{r}^{\,\prime}$};
    \node[squarednode](qc_inv)[right = 2cm of qc_out2]{Neural network};
   \node[roundnode] (rho_p)[ right= 2cm of qc_inv] {$\textbf{r}^{\,\prime \prime}$};
    \draw[->] (qc_out2.east) -- (qc_inv.west);
    \draw[->] (qc_inv.east) -- (rho_p);
    \end{tikzpicture}
\caption{A schematic representation of the quantum channel and the inversion process using a neural network. The quantum channel transforms the original Bloch vector $\textbf{r}$ into  $\textbf{r}^{\,\prime}$. The neural network predicts 
an approximation of the original Bloch vector denoted as $\textbf{r}^{\,\prime \prime}$.}
\label{inv_process}
\end{figure}
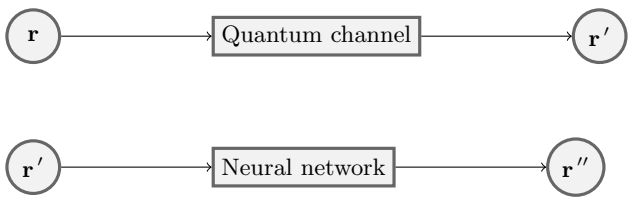

Let $\textbf{r}$ and $\textbf{r}^{\,\prime}$ represent the Bloch vectors of $\rho$ and $\mathcal{E}(\rho)$ respectively. Figure~\ref{inv_process} illustrates the inversion process we implemented. 

Our goal here is to minimize the difference between $\textbf{r}$ and  $\textbf r^{\, \prime \prime}$ while ensuring the process remains trace-preserving. Although mean squared error or fidelity as the loss function allows perfect recovery of the noise-free state, it increases the length of the Bloch vector{, thereby reducing its mixedness and making the network an unphysical operation since lost information cannot be recovered}. Therefore, in order to ensure that the neural network is a physically valid quasi-inverse, it should minimize the distance between its predicted noise-free Bloch vector $\textbf r^{\, \prime \prime}$ and the original noise-free vector $\textbf{r}$ without increasing the length of $\textbf r^{\, \prime \prime}$. For this purpose, we introduce the square of the modified trace distance (SMTD) as an error function. {This measure is newly introduced by us. The trace distance between two single-qubit mixed states $\rho$ and $\rho^{\, \prime \prime}$ is defined as~\cite{NC2010}
\begin{align}
    D(\rho^{\, \prime \prime}, \rho) &= \frac{1}{2}\text{tr}|\rho^{\, \prime \prime} - \rho | \\
    &= \frac{1}{4}\text{tr}|( \textbf{r}^{\, \prime \prime}- \textbf{r})\cdot \boldsymbol{\sigma}| \\
    &= \frac{1}{2}|( \textbf{r}^{\, \prime \prime}- \textbf{r})|,
\end{align}
where \textbf{r} and $\textbf{r}^{\prime\prime}$
are the Bloch vectors of $\rho$ and $\rho^{\, \prime \prime}$ respectively and $\pm |\textbf{r}^{\, \prime \prime}-\textbf{r}|$ are the eigenvalues of $(\textbf{r}^{\, \prime \prime}-\textbf{r} )\cdot\boldsymbol{\sigma}$.
Based on this definition, we introduce the modified trace distance
\begin{align}
    MTD &= \frac{1}{2}|( \textbf{r}^{\, \prime \prime}-r^{\prime}\textbf{r})|,
\end{align}
where $r^{\prime}$ is a scaling factor. Squaring it gives}
\begin{align}
    \text{SMTD} &= \frac{1}{4}[(r_{1}'' - r' r_{1})^{2} + (r_{2}'' - r' r_{2})^{2}+ (r_{3}'' - r' r_{3})^{2}],
\end{align}
where $r'= \sqrt{{r_{1}'}^{2}+{ r_{2}'}^{2}+{r_{3}'^{2}}}$ is the length of the  Bloch vector. We have introduced the scaling $r'$ so that the training of the neural network does not drive the neural network prediction $\textbf{r}^{\prime\,\prime}$ to match the original $\textbf{r}$ but rather match the scaled version $r'\textbf{r}$.

The loss function averaged over the entire dataset of size $M$ is the mean of the SMTD over the training dataset with $M$ examples:
\begin{equation}
    \text{MSMTD}  = \frac{1}{M}\sum_{i=1}^{M} \frac{1}{4}[(r_{i1}'' - r_i' r_{i1})^{2} + (r_{i2}'' - r_i' r_{i2})^{2}+ (r_{i3}'' - r_i' r_{i3})^{2}]\,,
\end{equation}
where $\textbf{r}_i$ is the Bloch vector of the ith example etc. 
The MSMTD measures the difference between the predicted and actual inverses, while the Adam optimizer updates model parameters during training. The model is trained for $100$ epochs and then used to predict the inverses of the test data.

\begin{figure}[h]
    \hspace*{0cm} % Adjust the value to move left
    \centering
    \begin{tikzpicture}
        \tikzstyle{unit}=[draw,shape=circle,minimum size=0.5cm, fill=gray!10]
		\tikzstyle{hidden}=[draw,shape=circle,minimum size=0.5cm, fill=gray!10]
        \tikzstyle{annot} = [text width=4em, text centered]
        \tikzstyle{annot_h} = [text width=10em, text centered]
        
        \foreach \i in {0,...,2} {
            \node[unit](x\i) at (2,\i){};
        }

        \foreach \i in {0,...,2}{
            \node[hidden](h\i) at (4.5,\i +0.7){};
        }

        \node at (4.5,0.1){\vdots};
		\node[hidden](h3) at (4.5,-0.7){};
        \foreach \i in {0,...,3} {
            \foreach \j in {0,...,2} {
                \draw[->] (x\j) -- (h\i);
            }
        }
        \foreach \i in {0,...,2} {
            \node[unit](y\i) at (7,\i){};
        }
        \foreach \i in {0,...,3} {
            \foreach \j in {0,...,2} {
                \draw[->] (h\i) -- (y\j);
            }
        }
 
        \node[annot,above of=x2, node distance=1cm] (input) {Input layer};
        \node[annot,above of=h2, node distance=1cm] (dense) {Dense layer};
        \node[annot,above of=y2, node distance=1cm] (output) {Output layer};

        %\node[annot,below of=h3, node distance=1cm] (dense_nns);

    \end{tikzpicture}

\caption{A schematic representation of our fully connected neural network.}
\label{nn_arch}
\end{figure}
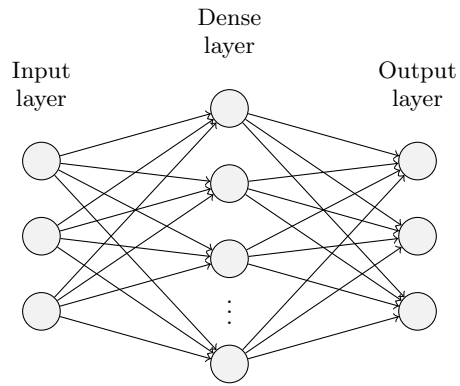

\section{Training and evaluation}\label{training}

To generate random density matrices in Python, we define a function that generates random matrices using NumPy and then converts them into valid density matrices. For our model, we generate $1000$~$n \times n$ density matrices and then apply a quantum channel to them. The process of applying a quantum channel can also be achieved by defining a function that takes in $\rho$ and outputs $\mathcal{E}(\rho)$ after applying the desired quantum channel to it. After generating $\rho$ and $\mathcal{E}(\rho)$, their Bloch vector components are computed using $r_{i} = \text{Tr}[\rho\sigma_{i}]$. We used $90$\% of the samples for training and $10$\% for testing the model.
The Bloch vectors of the outputs of the channel $\mathcal{E}(\rho)$ and those of the density matrices $\rho$ are fed into the network for training. $\textbf r^{\, \prime}$  is fed as the input while $\textbf r$ is fed as the output, allowing the network to learn an inverse mapping between the two states.

The neural network model is designed as a simple feedforward network with fully connected layers. In our model, we first add an Input layer with $3$ neurons corresponding to the three components of the Bloch vector. Then, we add a Dense layer with $32$ neurons. Finally, we add an Output layer with $3$ neurons corresponding to the $3$ components of the predicted Bloch vector. {Only linear activation functions are used to ensure that the network behaves as a linear operation.} Figure~\ref{nn_arch} shows the layout of our neural network. 

\begin{figure}
  \centering  {\includegraphics[width=0.485\textwidth]{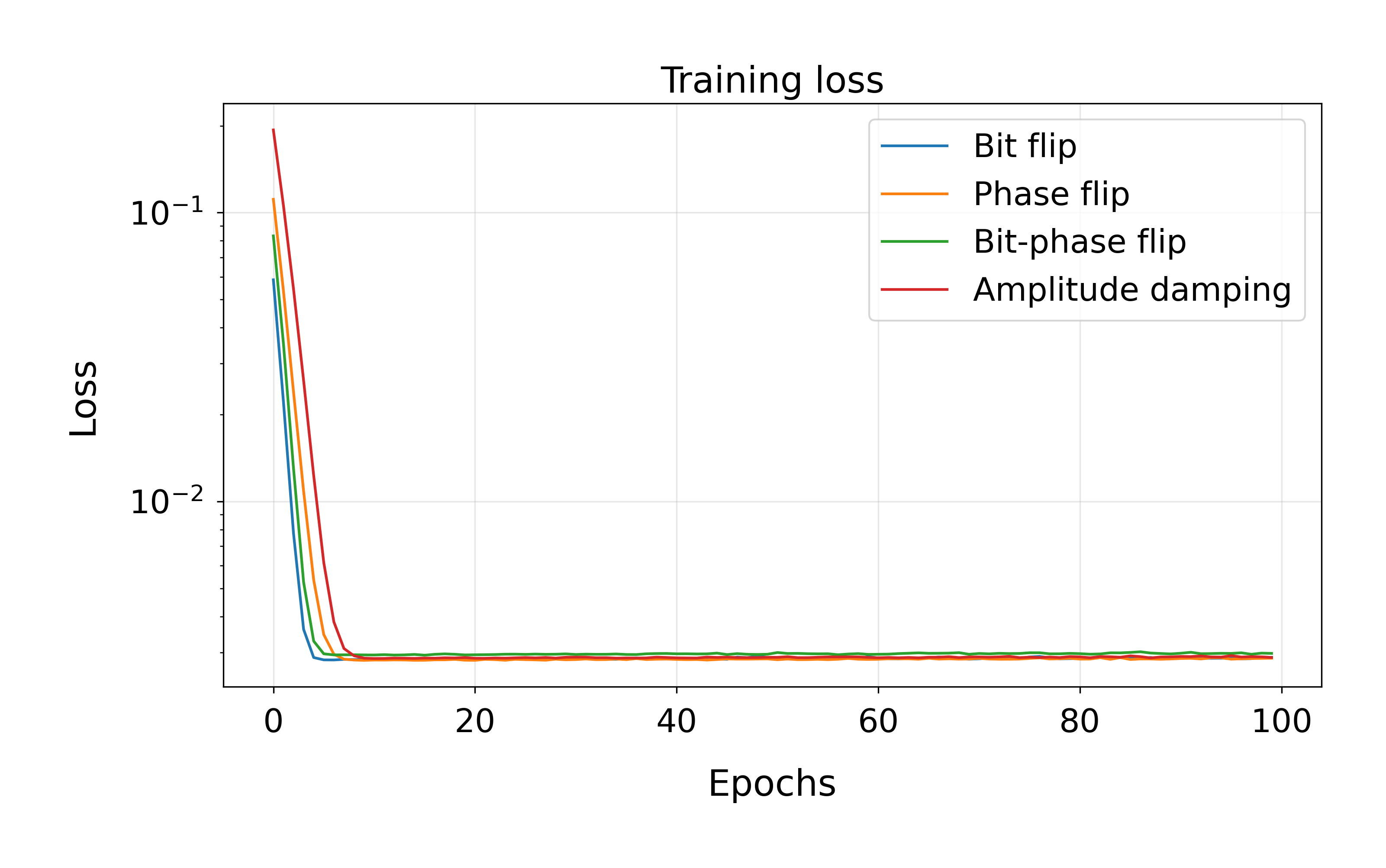}}
  \caption{Training loss history of neural networks trained to invert quantum channels. }
  \label{tr_loss}
\end{figure}

We trained and tested our model for the bit flip, phase flip, bit-phase flip, and amplitude damping channels. Figure~\ref{tr_loss} shows the training loss across different quantum channels.

\begin{figure*}[h]
    \centering
    %---- (a) ----
    \begin{subfigure}[b]{0.47\textwidth}
        \centering
        \includegraphics[width=\textwidth]{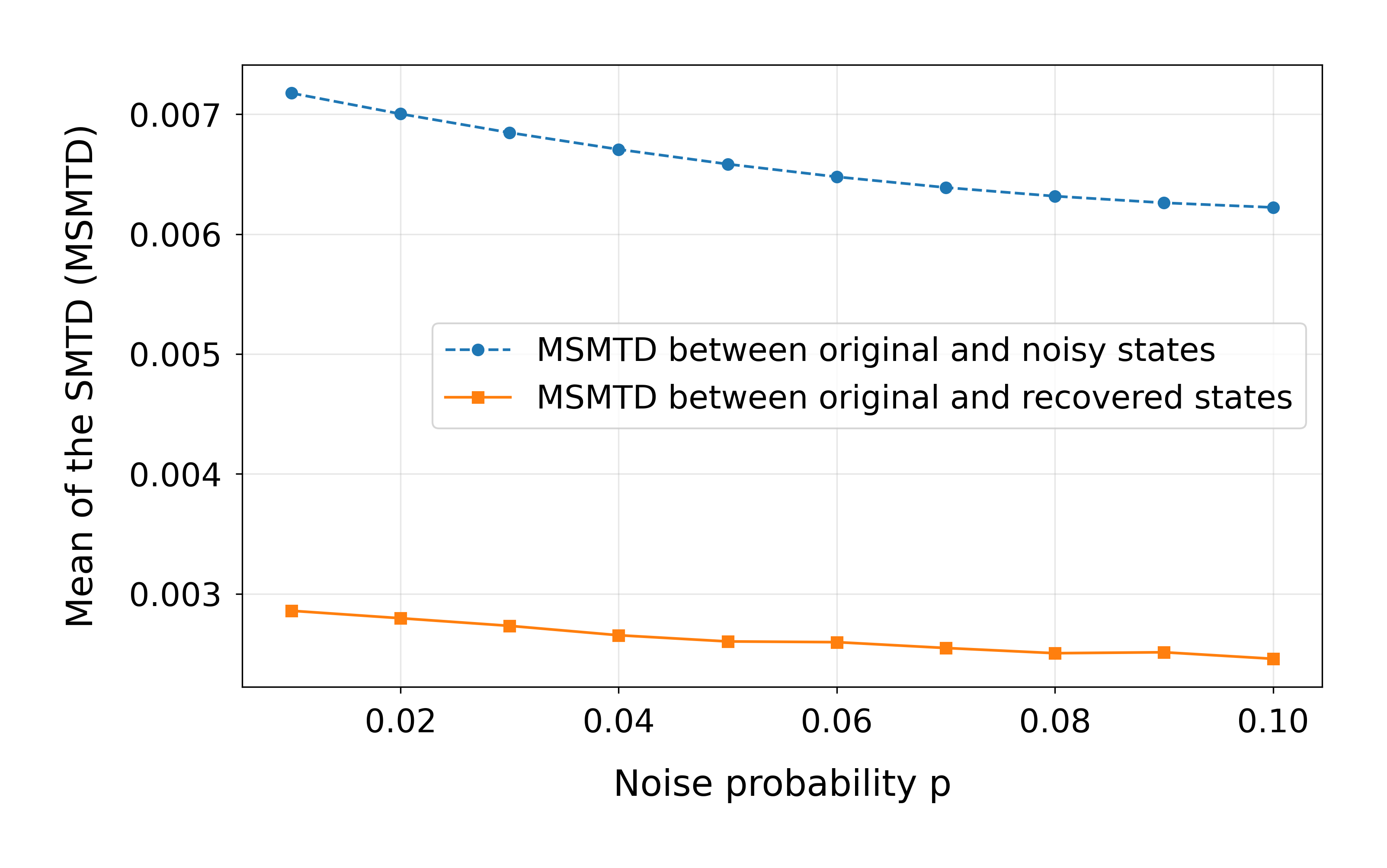}
        \caption{}
        \label{fig:bit}
    \end{subfigure}
    \begin{subfigure}[b]{0.47\textwidth}
        \centering
        \includegraphics[width=\textwidth]{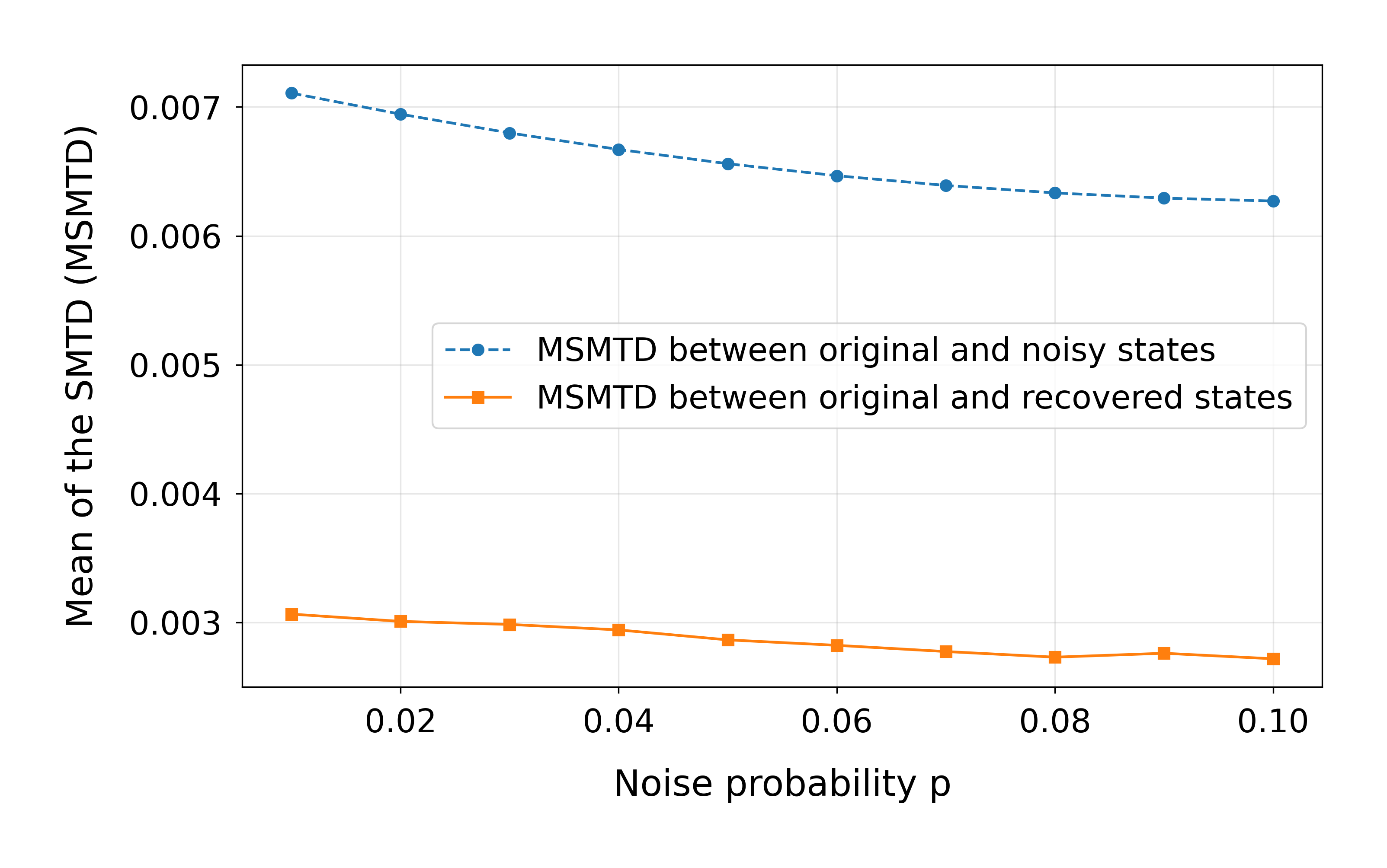}
        \caption{}
        \label{fig:phase}
    \end{subfigure}
    \begin{subfigure}[b]{0.47\textwidth}
        \centering
        \includegraphics[width=\textwidth]{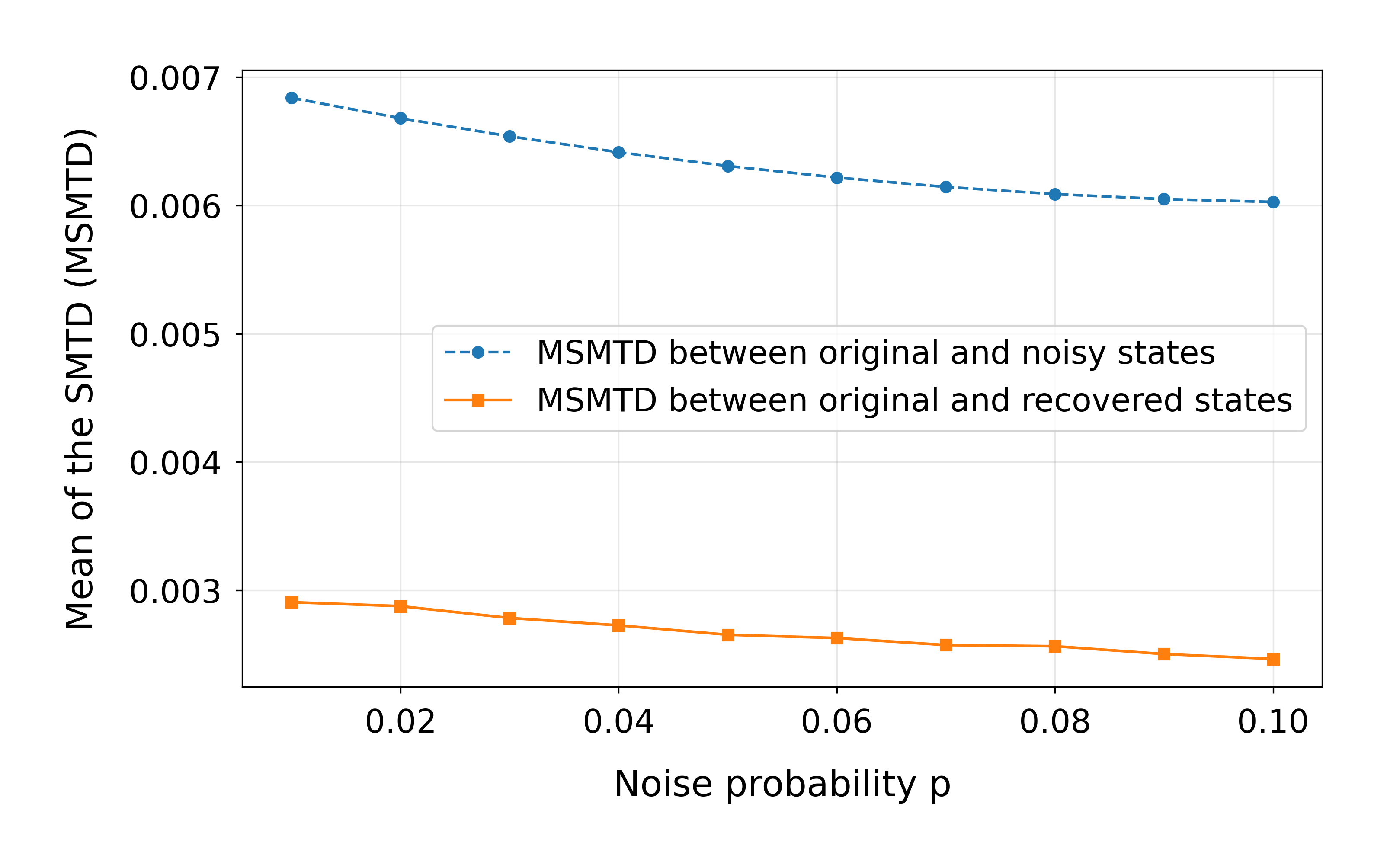}
        \caption{}
        \label{fig:bit-phase}
    \end{subfigure}
    \begin{subfigure}[b]{0.47\textwidth}
        \centering
        \includegraphics[width=\textwidth]{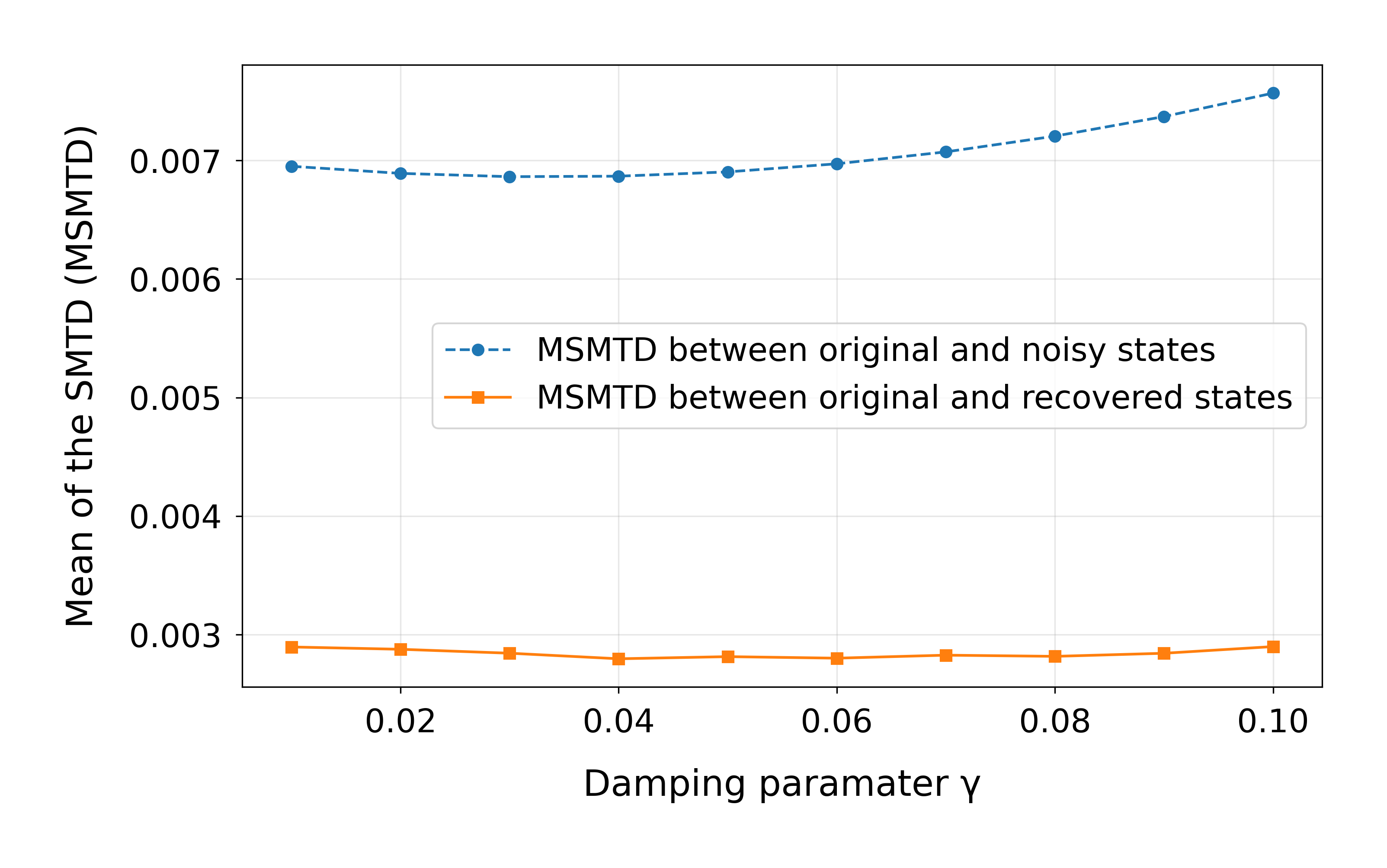}
        \caption{}
        \label{fig:ad}
    \end{subfigure}

    \caption{{Mean of the square of the modified trace distance (MSMTD) of single qubit (a) bit flip, (b) phase flip and (c) bit-phase flip channels for varying noise probability $p$ and (d) amplitude damping channel for varying damping parameter $\gamma$.}}
    \label{fig:msmtd1}
\end{figure*}

{The neural network was then trained for different values of $p$ and $\gamma$ and the SMTD was computed between the original and noisy states as well as the original and recovered states for a dataset of $1000$ samples. The mean values are plotted in Figure~\ref{fig:msmtd1}.}

\section{Quantum process tomography}\label{tomography}
Quantum process tomography is a technique used to characterize an unknown quantum process. The goal is to find the Kraus operators $E_{i}$ of the unknown process $\mathcal{E}(\rho)$. For a $d$-dimensional system, we start by choosing $d^{2}$ fixed operators {$\tilde{E}_{m}$} which form an operator basis,
\begin{equation}
    E_{i} = \sum \limits_{m} e_{im} \tilde{E}_{m},
\end{equation}
for some set of complex numbers $e_{im}$. Thus, we can write $\mathcal{E}(\rho)$ as
\begin{equation}\label{chi_e}
    \mathcal{E}(\rho) = \sum \limits_{m,n} \tilde{E}_{m} \rho \tilde{E}_{n}^{\dag} \chi_{mn},
\end{equation}
where $\chi_{mn} \equiv \sum 
\limits_{i} e_{im}e_{in}^{*}$ is a positive, Hermitian matrix known as the chi matrix or the process matrix, which contains enough information to completely describe $\mathcal{E}(\rho)$ once the operators $E_{i}$ have been fixed. The goal, therefore, is to find the $\chi$ matrix, which can be accomplished following the scheme outlined in{~\cite{Chuang01111997}}.

We choose $d^{2}$ pure quantum states such that their density matrices form a basis $\left\{ \rho_{j} \right\}_{j=1} ^{d^{2}}$ for the space of $d\times d$ matrices. A common choice is the set of pure states $\{|n\rangle, |m\rangle, |+\rangle = \frac{|n\rangle+|m\rangle}{\sqrt{2}}, |-\rangle=\frac{|n\rangle+i|m\rangle}{\sqrt{2}}\} $ since any $\mathcal{E}(|n\rangle \langle m|)$ can be obtained by forming linear combinations of $\mathcal{E}(|n\rangle \langle n|)$, $\mathcal{E}(|m\rangle \langle m|)$, $\mathcal{E}(|+\rangle \langle +|)$, $\mathcal{E}(|-\rangle \langle -|)$.

Each $\mathcal{E}(\rho_{j})$ is a linear combination of the basis states,
\begin{equation}\label{lambda_e}
    \mathcal{E}(\rho_{j}) = \sum \limits_{k} \lambda_{jk} \rho_{k},
\end{equation}
where $\lambda_{jk}$ may be determined through algebraic manipulation. Furthermore, we write
\begin{equation}\label{beta_e}
    \tilde{E}_{m} \rho \tilde{E}_{n}^{\dag} = \sum 
    \limits_{k}\beta_{jk}^{mn} \rho_{k},
\end{equation}
where $\beta_{jk}^{mn}$ are complex numbers which can be easily determined using $\tilde{E_{i}}$ and $\rho_{j}$. Now inserting Eq.~(\ref{lambda_e}) and Eq.~(\ref{beta_e}) into Eq.~(\ref{chi_e}), we get
\begin{equation}
    \sum \limits_{k}\sum \limits_{mn}\chi_{mn} \beta_{jk}^{mn} \rho_{k} = \sum \limits_{k}\lambda_{jk}\rho_{k},
\end{equation}
which can be written as
\begin{equation}
    \sum \limits_{mn} \beta_{jk}^{mn}\chi_{mn}  = \lambda_{jk},
\end{equation}
for each $k$.

We interpret $\lambda$ and $\chi$  as column vectors, and $\beta$ as a $d^{4}\times d^{4}$ matrix with columns indexed by $mn$ and rows by $jk$. Then, the problem merely comes down to finding an inverse of $\beta$ and multiplying it with $\lambda$ to get $\chi$ and then reshaping $\chi$,
\begin{equation}\label{final_chi}
    \chi = \beta^{-1} \lambda.
\end{equation}

The $\chi$ matrix may then be used to find the Kraus operators $E_{i}$~\cite{NC2010}. Let the unitary matrix $U^{\dag}$ diagonalize $\chi$,
\begin{equation}
    \chi_{mn} =  \sum \limits_{xy}U_{mx}d_{x}\delta_{xy}U^{*}_{ny}.
\end{equation}

Then, the Kraus operators can be found using
\begin{equation}
    E_{i} = \sqrt{d_{i}}\sum \limits_{j}U_{ji}E_{j}.
\end{equation}

In our case, we wish to find the $\chi$ 
matrix to identify the process through which our neural network is inverting the quantum channel outputs. Following the procedure explained above, we first choose a set of input states. We choose
\begin{equation*}
    \left\{ \rho_{j} \right\}_{j=1} ^{d^{2}} = \left\{|0\rangle \langle 0|, |1\rangle \langle 1|, |+\rangle \langle +|, |-\rangle \langle -| \right\}.
\end{equation*}

These $\rho_{j}$ serve as inputs for our neural network, producing outputs of the form $\mathcal{E}(\rho_{j})$. Making use of Eq.~(\ref{lambda_e}), we see that $\lambda_{jk}$ can be obtained using
\begin{equation}
    \lambda_{jk} = tr(\mathcal{E}(\rho_{j})^{\dag}\rho_{k}).
\end{equation}

Then, we choose $d\times d$ Pauli matrices as our set of operators $\left\{\tilde{E}_{i}\right\}_{i=1}^{d^{2}}$. Then, using Eq.~(\ref{beta_e}), we see that the values of the $\beta$ matrix can be found in the following manner
\begin{equation}
    \beta_{jk}^{mn} = tr(\rho_{k}^{\dag} \tilde{E}_{m}\rho_{j}\tilde{E}_{n}^{\dag}).
\end{equation}

Now that we have $\lambda$ and $\beta$, we use Eq.~(\ref{final_chi}) to find $\chi$. {The results for our desired channels can be found in~\ref{appendix1}-\ref{appendix4}. Analytically, we find the quasi-inverses for our targeted $p$ and $\gamma$ values to be the identity channel which offers no decrease in the SMTD. Our neural network, when tested on a sample of 1000 randomly generated density matrices, reduces the SMTD in around $80\%$ cases. Analysis of the remaining $20\%$ states shows that their average purity is around $0.97$ and for such states, we see that at small noise probabilities, the SMTD between the original and noisy states is already very small and any further reduction in the Bloch vector length causes the SMTD between the original and predicted states to increase.}

\section{Conclusions}\label{conc}
We have demonstrated the use of classical neural networks to approximate the quasi-inverse by leveraging the Bloch vector representation of quantum states. The model effectively reversed the effect of noise introduced by various quantum channels. We used quantum process tomography to characterize the quasi-inversion process under different noise conditions. Also, it was found that the chosen loss function trained the neural network to be a completely positive and trace-preserving (CPTP) map which can be realized physically. Future research could focus on the use of these quasi inverses for noise mittigiation in quantum computing.

\appendix 
\section{Bit flip channel} \label{appendix1}
Following this procedure for the bit flip channel with $p=0.01$, we obtain the following chi matrix representing the neural network's quasi-inverse
\begin{equation*}
   \chi = 
   \begin{pmatrix}
    0.8766 & 0 & 0 & 0\\
    0 & 0.0405 & 0 & 0\\
    0 & 0 & 0.0402 & 0\\
    0 & 0 & 0 & 0.0426\\
   \end{pmatrix}.
\end{equation*}\\

Using this $\chi$ matrix, we find the Kraus operators, which turn out to be
\begin{equation*}
    E_{0} = 0.2006\sigma_{2},
\end{equation*}
\begin{equation*}
    E_{1} = 0.2012\sigma_{1},
\end{equation*}
\begin{equation*}
    E_{2} = 0.2064\sigma_{3},
\end{equation*}
\begin{equation*}
    E_{3} = 0.9363I.
\end{equation*}

For $p=0.05$, we obtain the following $\chi$ matrix:
\begin{equation*}
    \chi = 
    \begin{pmatrix}
    0.8754 & 0 & 0 & 0\\
    0 & 0.0215 & 0 & 0\\
    0 & 0 &0.0512 & -0.0017\\
    0 & 0 & -0.0017 & 0.0521\\
    \end{pmatrix}.
\end{equation*}
The corresponding Kraus operators are
\begin{equation*}
    E_{0} = 0.1466\sigma_{1},
\end{equation*}
\begin{equation*}
    E_{1} = -0.1770\sigma_{2}-0.1362\sigma_{3},
\end{equation*}
\begin{equation*}
    E_{2} = -0.1410\sigma_{2}+0.1831\sigma_{3},
\end{equation*}
\begin{equation*}
    E_{3} = 0.9356I.
\end{equation*}
\section{Phase flip channel}\label{appendix2}
For the phase flip channel with $p=0.01$, we get
\begin{equation*}
   \chi = 
   \begin{pmatrix}
    0.8825 & 0 & 0 & 0\\
    0 & 0.0420 & 0 & 0\\
    0 & 0 & 0.0402 & 0.0015\\
    0 & 0 & 0.0015 & 0.0354\\
   \end{pmatrix}.
\end{equation*}

Using this $\chi$ matrix, we find the Kraus operators, which turn out to be
\begin{equation*}
    E_{0} = 0.0516\sigma_{2}-0.1798\sigma_{3},
\end{equation*}
\begin{equation*}
    E_{1} = -0.1938\sigma_{2}-0.0556\sigma_{3},
\end{equation*}
\begin{equation*}
    E_{2} = 0.2049\sigma_{1},
\end{equation*}
\begin{equation*}
    E_{3} = 0.9394I.
\end{equation*}

For $p=0.05$, we get
\begin{equation*}
    \chi = 
   \begin{pmatrix}
    0.8794 & 0 & 0 & 0\\
    0 & 0.0507 & 0.0015 & 0.0028\\
    0 & 0.0015 & 0.0484 & 0\\
    0 & 0.0028 & 0 & 0.0217\\
   \end{pmatrix}.
\end{equation*}
The corresponding Kraus operators are
\begin{equation*}
    E_{0} = 0.0140\sigma_{1}-0.0008\sigma_{2}-0.1457\sigma_{3},
\end{equation*}
\begin{equation*}
    E_{1} = 0.0906\sigma_{1}-0.1985\sigma_{2}+0.0098\sigma_{3},
\end{equation*}
\begin{equation*}
    E_{2} = 0.2057\sigma_{1}+0.0948\sigma_{2}+0.0192\sigma_{3},
\end{equation*}
\begin{equation*}
    E_{3} = 0.9378I.
\end{equation*}
\section{Bit-phase flip channel}\label{appendix3}
For the bit-phase flip channel with $p=0.01$, we get
\begin{equation*}
   \chi = 
   \begin{pmatrix}
    0.8736 & 0 & 0 & 0\\
    0 & 0.0459 & -0.0014 & 0\\
    0 & -0.0014 & 0.0386 & 0\\
    0 & 0 & 0 & 0.0419
   \end{pmatrix}.
\end{equation*}
The corresponding Kraus operators are
\begin{equation*}
    E_{0} = -0.0357\sigma_{1}-0.1925\sigma_{2},
\end{equation*}
\begin{equation*}
    E_{1} = 0.2047\sigma_{3},
\end{equation*}
\begin{equation*}
    E_{2} = -0.2112\sigma_{1}+0.0391\sigma_{2},
\end{equation*}
\begin{equation*}
    E_{3} = 0.9347I.
\end{equation*}

For $p=0.05$, we get
\begin{equation*}
    \chi = 
   \begin{pmatrix}
    0.8761 & 0 & 0 & 0\\
    0 & 0.0507 & 0 & -0.0018\\
    0 & 0 & 0.0228 & 0\\
    0 & -0.0018 & 0 & 0.0504\\
   \end{pmatrix}.
\end{equation*}
The corresponding Kraus operators are
\begin{equation*}
    E_{0} = 0.1510\sigma_{2},
\end{equation*}
\begin{equation*}
    E_{1} = -0.1495\sigma_{1}-0.1625\sigma_{3},
\end{equation*}
\begin{equation*}
    E_{2} = -0.1684\sigma_{1}+0.1549\sigma_{3},
\end{equation*}
\begin{equation*}
    E_{3} = 0.9360I.
\end{equation*}
\section{Amplitude damping channel}\label{appendix4}
For the amplitude damping channel with $\gamma = 0.01$, we get
\begin{equation*}
    \chi = 
   \begin{pmatrix}
    0.8736 & 0 & 0 & 0\\
    0 & 0.0404 & 0.0043 & -0.0036\\
    0 & 0.0043 & 0.0444 & -0.0017\\
    0 & -0.0036 & -0.0017 & 0.0416\\
    \end{pmatrix}.
\end{equation*}
The corresponding Kraus operators are
\begin{equation*}
    E_{0} = 0.1559\sigma_{1}-0.0661\sigma_{2}+0.0886\sigma_{3},
\end{equation*}
\begin{equation*}
    E_{1} = 0.0355\sigma_{1}-0.1248\sigma_{2}-0.1556\sigma_{3},
\end{equation*}
\begin{equation*}
    E_{2} = 0.1218\sigma_{1}+0.1564\sigma_{2}-0.0977\sigma_{3},
\end{equation*}
\begin{equation*}
    E_{3} = 0.9347I.
\end{equation*}

For $\gamma = 0.05$, we get
\begin{equation*}
    \chi = 
   \begin{pmatrix}
    0.8770 & 0 & 0 & 0\\
    0 &0.0399 & 0.0019 & 0\\
    0 & 0.0019 & 0.0399 & 0.0036\\
    0 & 0 & 0.0036 & 0.0432\\
    \end{pmatrix}.
\end{equation*}
The corresponding Kraus operators are
\begin{equation*}
    E_{0} = -0.0893\sigma_{1}+0.1480\sigma_{2}-0.0826\sigma_{3},
\end{equation*}
\begin{equation*}
    E_{1} = 0.1744\sigma_{1}+0.0597\sigma_{2}-0.0812\sigma_{3},
\end{equation*}
\begin{equation*}
    E_{2} = 0.0397\sigma_{1}+0.1199\sigma_{2}+0.1725\sigma_{3},
\end{equation*}
\begin{equation*}
    E_{3} = 0.9347I.
\end{equation*}

\section{CPTP check}
For all the examples considered in this paper, we checked that $\sum\limits_{i} E^{\dag}_{i} E_{i} = I+\delta$ {where $\delta\leq10^{-15}$ for $0\leq p\leq 0.1$}, validating that the neural network is a completely positive and trace-preserving (CPTP) quantum channel.

\bibliographystyle{apsrev4-2}
\bibliography{apssamp}% Produces the bibliography via BibTeX.
\end{document}